\documentclass[preprint]{vgtc}               

\graphicspath{{figures/}{pictures/}{images/}{./}} 

\usepackage{times}                     

\usepackage{tabu}                      
\usepackage{booktabs}                  
\usepackage{lipsum}                    
\usepackage{mwe}                       

\usepackage{mathptmx}                  
\usepackage{tcolorbox}
\usepackage{xcolor}
\usepackage{listings}
\usepackage{enumitem}
\usepackage[table]{xcolor}
\usepackage[normalem]{ulem}
\usepackage{amsmath,amsfonts}
\usepackage{graphicx}  
\usepackage{array, multirow, booktabs}
\usepackage[nocompress, sort]{cite}

\definecolor{result_red}{HTML}{FFA9AD}
\definecolor{result_orange}{HTML}{FFD5A8}
\definecolor{result_green}{HTML}{9CD9B6}

\definecolor{lightred}{RGB}{255, 175, 177}
\definecolor{lightorange}{RGB}{255, 215, 174}
\definecolor{lightyellow}{RGB}{255, 250, 194}

\definecolor{heatmap_purple}{HTML}{6320D2}
\definecolor{heatmap_yellow}{HTML}{F6D746}

\definecolor{boxplot_gray}{HTML}{B1B1B1}
\definecolor{boxplot_green}{HTML}{A8DEAF}
\definecolor{boxplot_red}{HTML}{F497A3}

\definecolor{kim_green}{HTML}{4CBB17}
\definecolor{swapnil_purple}{HTML}{9300FF}
\definecolor{ari_red}{HTML}{EC1313}

\newcommand{\para}[1]{\vspace{0.35em}\noindent\normalsize\textbf{#1.}\xspace}
\newcommand{\parac}[1]{\vspace{0.35em}\noindent\normalsize\textbf{#1:}\xspace}

\newcommand{\pre}{Pre-calibration session\xspace}
\newcommand{\post}{Post-calibration session\xspace}

\newcommand{\hori}{$|\Delta{\theta}|$}
\newcommand{\verti}{$|\Delta{\phi}|$}
\newcommand{\dist}{$\alpha$}

\onlineid{1022}

\vgtccategory{Research}

\vgtcinsertpkg

\usepackage{orcidlink}

\preprinttext{%
  \ifodd\value{page}%
    \parbox{0.9\textwidth}{%
      © 2026 IEEE. This is the author's version of the article that appeared at the IEEE Conference on Virtual Reality and 3D User Interfaces Abstracts and Workshops (IEEE VRW). The final version of this record is available at: \href{https://doi.org/10.1109/VRW70859.2026.00168}{\textcolor{blue}{\textit{10.1109/VRW70859.2026.00168}}}
    }%
  \fi
}

\title{Evaluating Spatialized Auditory Cues for Rapid Attention Capture in XR}

\author{Yoonsang Kim\thanks{e-mail:yoonsakim@cs.stonybrook.edu} %
\and Swapnil Dey\thanks{e-mail:swdey@cs.stonybrook.edu} %
\and Arie E. Kaufman\thanks{e-mail:ari@cs.stonybrook.edu}}
\affiliation{\scriptsize Center for Visual Computing, Stony Brook University}

\teaser{
  \centering
  \includegraphics[width=\linewidth]{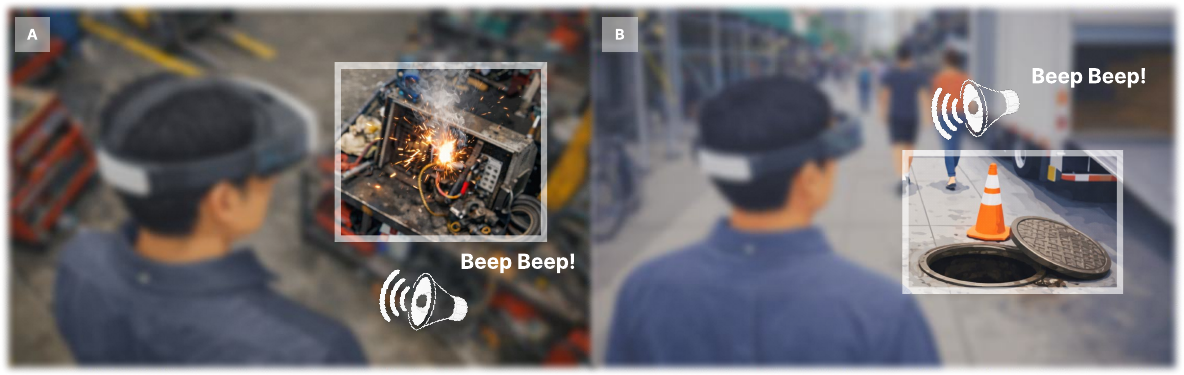}
  \caption{Conceptual illustration of spatial audio as an immediate attention-guidance mechanism in XR. In both scenarios, a brief spatialized auditory cue (``Beep Beep'') provides coarse directional information that rapidly orients the user's attention toward a target without any visual guidance. Example use-cases include (A) industrial XR maintenance, where hazards or points of interest must be recognized quickly, and (B) outdoor wayfinding, where obstacles outside the user's current focus, require rapid attention redirection.}
  \label{fig:teaser}
}

\abstract{
In time-critical eXtended reality (XR) scenarios where users must rapidly reorient their attention to hazards, alerts, or instructions while engaged in a primary task, spatial audio can provide an immediate directional cue without occupying visual bandwidth. However, such scenarios can afford only a brief auditory exposure, requiring users to interpret sound direction quickly and without extended listening or head-driven refinement. This paper reports a controlled exploratory study of rapid spatial-audio localization in XR. Using HRTF-rendered broadband stimuli presented from a semi-dense set of directions around the listener, we quantify how accurately users can infer coarse direction from brief audio alone. We further examine the effects of short-term visuo-auditory feedback training as a lightweight calibration mechanism. Our findings show that brief spatial cues can convey coarse directional information, and that even short calibration can improve users' perception of aural signals. While these results highlight the potential of spatial audio for rapid attention guidance, they also show that auditory cues alone may not provide sufficient precision for complex or high-stakes tasks, and that spatial audio may be most effective when complemented by other sensory modalities or visual cues, without relying on head-driven refinement. We leverage this study on spatial audio as a preliminary investigation into a first-stage attention-guidance channel for wearable XR (e.g., VR head-mounted displays and AR smart glasses), and provide design insights on stimulus selection and calibration for time-critical use.} 


\begin{document}


\firstsection{Introduction} 
\maketitle


Spatial audio has been explored in eXtended Reality (XR) systems as a means to guide user attention and convey spatial information beyond the limited Field-of-View (FoV) of wearable displays. For Head-Mounted Displays (HMD) and smart glasses, where both the physical FoV and the user's on-screen visual resources (e.g., overlays competing for attention) are constrained, spatialized sound can provide a complementary channel for directing attention without adding additional visual elements to the display. Spatial audio can indicate the approximate direction of off-screen or off-view targets, serving as an initial cue that enables users to reorient their view before more precise visual guidance becomes available~\cite{cho2025evaluating, binetti2021using}.

The human ability to infer sound direction relies on a combination of binaural and monaural auditory cues. These enable users to infer coarse spatial direction even in the absence of visual input, making spatial audio suitable for attention guidance in visually constrained environments~\cite{huang1998spatial, carliniaudio}. Moreover, prior studies have shown that auditory spatial perception is subject to systematic ambiguities, and that other factors such as sound bandwidth, stimulus duration, head movement~\cite{wallach1940role, houtgast1994stimulus}, reverberation, and perceptual calibration through training~\cite{king2009visual}, can further reduce the ambiguity and spatial audio localization uncertainty~\cite{carliniaudio, cho2024auptimize, rajguruspatial}. Despite the usefulness of spatial audio for attention guidance, there is limited empirical understanding of how accurately users can interpret spatialized auditory cues under immediate, time-constrained conditions. In particular, there is a research gap in how stimulus characteristics and short-term perceptual calibration influence users' ability to extract coarse directional information from short auditory exposures. 

To this end, we conduct a controlled exploratory study (N=17) that characterizes rapid spatial-audio localization under two factors: (1) stimulus emission direction around the listener and (2) the presence or absence of short visuo-auditory feedback training. We frame spatial audio not as a mechanism for achieving precise localization through extended exploration~\cite{macpherson2000localization}, but as an immediate attention-guidance cue that supports fast orienting responses, even with brief exposure. This framing is motivated by XR application scenarios such as industrial hazard notification and obstacle avoidance in outdoor wayfinding, where users must quickly interpret spatial cues within and outside FoV (when the user's attention was momentarily distributed), before any assistance via a visual cue. 

By systematically quantifying localization performance under these conditions, our findings ground the use of spatial audio as an immediate attention cue in wearable XR. We clarify the perceptual boundaries of rapid auditory interpretation, and share design insights for XR systems that may leverage spatial audio for time-critical interaction, as well as for audio-based AI assistant systems in which spoken references or AI-generated notifications must efficiently direct user attention to spatial targets. Based on the above motivations, we investigate the following Research Questions (RQs) in this paper:

\begin{enumerate}[label=\textbf{RQ\arabic*.}, leftmargin=*, align=left]
\item How accurately can users localize spatialized auditory cues across azimuth and elevation under an immediate, time-constrained condition, without relying on extended exploration time or head-driven refinement?
\item How does localization accuracy vary across coarse directional regions (front, back, left, right, up and down) under short auditory exposure?
\item Does a short-term feedback training increase immediate spatial-audio localization accuracy and shape user confidence during post-training judgments?
\end{enumerate}


\section{Related Work} 
\subsection{Multi-sensory Notification in XR} 
Multi-sensory notification techniques for directing user attention leverage visual, auditory, and tactile modalities to support awareness and task performance in XR systems. These modalities are commonly used to compensate for limited visual bandwidth, divided attention, and the spatial separation between a user's current focus and relevant events or targets. Prior works have shown that visual notifications are particularly effective for targets within the user's current FoV, where overlays, highlights, or visual markers can convey precise spatial information with low ambiguity when visual resources are available~\cite{cho2025evaluating, lin2021labeling, petford2019comparison, choi2025distance, gruenefeld2017visualizing, pluisch2025extended}. When targets lie outside the user's view or when visual bandwidth is constrained, non-visual modalities become increasingly important. Auditory notifications have been shown to improve awareness and response performance for out-of-view targets by providing directional cues that prompt users to reorient their attention~\cite{binetti2021using, hinzmann2025finding, feng2023can, yang2020effects, bhattacharyya2025birds, bak2025beyond}. Comparative studies further indicate that audio-based cues can outperform or complement visual-only indicators for out-of-view notification, particularly when persistent visual elements would interfere with the primary task~\cite{petford2019comparison, marquardt2020comparing}.

Haptic notifications have also been explored as a complementary modality for attention guidance in XR, especially when visual and auditory channels are heavily loaded. Prior work shows that vibro-tactile cues can support alerting and attention shifts, but their effectiveness depends on timing, task context, and how they are combined with other modalities~\cite{trepkowski2021multisensory, kaur2025senses, trapero}. Beyond individual modality effectiveness, a study on modality congruence emphasizes that interactions between sensory channels play a critical role in notification design, showing that specific types of task-notification modality pairings may introduce interference, increasing cognitive load and reducing performance~\cite{kaur2025senses}. As a result, tactile cues are used to reinforce visual or auditory signals rather than to convey precise spatial information.

The literature provides guidance on how different sensory modalities can support attention in XR. However, there remains a research gap in understanding spatial attention under immediate, time-constrained conditions, where sustained exposure, exploration, or head movement~\cite{wallach1940role, houtgast1994stimulus} is not ideal. In this work, we examine how auditory cues are interpreted when used as immediate attention signals, focusing on the perceptual performance of spatial audio for rapid attention guidance in XR.

\subsection{Aural Perception and Cues}
Aural perception enables rapid attention direction by allowing listeners to infer the spatial direction of sound sources without requiring visual engagement. Spatial hearing relies on multiple auditory cues that are differentially informative across frequency ranges and spatial dimensions. Interaural Time Differences (ITD) constitute the dominant cue for horizontal (azimuth) localization for sounds below 1,500 Hz, due to the relationship between the wavelength of low-frequency sound and the physical spacing between the ears~\cite{rayleigh1907xii, stevens1936localization}. Human perceptual system uses ITD a temporal differences of sound arrival at the two ears for azimuth estimation, to estimate the azimuthal position of a sound source. For frequencies above 1,500 to 2,000 Hz, the Interaural Level Differences (ILD) become more informative. Furthermore, horizontal localization performance degrades when stimuli contain only frequencies below approximately 2,000 Hz or only above approximately 12,000 Hz, indicating that the dominant contributions to horizontal localization arise from a mid-band frequency range~\cite{morikawa2010signal}. The anatomical features of a human head create acoustic shadowing that produces asymmetries at higher frequencies. When broadband stimuli span both low and high frequencies, the two binaural cues can lower the estimated horizontal localization error~\cite{carliniaudio}. The vertical (elevation) localization and front-back audio source identification primarily rely on monaural spectral cues transformed by the human anatomy such as the shape of the outer ear (pinna), head, and torso. These structures introduce direction-dependent filtering that produces spectral peaks and notches at high frequencies, approximately above 4,000 Hz, with the most pronounced cues in the 6,000 to 9,000 Hz band~\cite{zonooz, blauert1997spatial, pintoperception, yao2020role, andeolspatial}. On the other hand, Asano et al. claim that elevation judgments rely on cues in the high-frequency region above 5,000 Hz~\cite{asano1990role}.

To combine the complementary localization benefits of each auditory cue within a single stimulus, we use filtered Gaussian noise~\cite{valzolgher2020impact, cho2024auptimize} as our broadband tone. Our stimulus spans the low-band that provides robust ITD cues (below 1,500 Hz), the mid-band where ILD grows with head-shadowing effects (above 1,500 Hz), and the high-band region where spectral notches and peaks encode elevation cues (6,000 to 9,000 Hz). This wide-band ($>1$ octave wide) stimulus leverages the benefits of all cues simultaneously, producing improved sound source localization accuracy~\cite{yost2014sound}.

\subsection{Spatial Audio in XR}
Spatial audio can be applied to XR systems using Head-Related Transfer Functions (HRTFs). HRTFs approximate how sound propagates from a source to a listener's ears. HRTFs encode the combined effects of ITD, ILD, and spectral cues, enabling headphone-based reproduction of a spatialized sound source. The anatomy of the listener differs by person, however, a generic HRTF that generalizes the transformation of the audio signal, are generally known to be sufficient (to an extent) for conveying coarse directional information~\cite{hrtfisgoodenough, carliniaudio, boem2025spatial}. This makes spatial audio more suitable as a situated attention-directing mechanism than a precise localization tool. Studies have shown that object-anchored audio supports user-awareness and task performance in scenarios involving out-of-view targets, navigation, and remote collaboration, particularly when visual bandwidth is limited~\cite{binetti2021using, hinzmann2025finding, arce2017effects}. In addition, a recent work by Cho et al. leverages the psycho-acoustic properties of spatial audio, where humans are unable to precisely pinpoint an audio source due to the cone-of-confusion~\cite{cho2024auptimize}.

Motivated by these works, we leverage spatial audio as a rapid, coarse attention-guidance mechanism for time-sensitive (immediate) use cases such as industrial hazard notification and outdoor obstacle avoidance assistance.

\subsection{Sensory Calibration and Learning}
While the combined use of binaural cues and monaural cues, provides relatively robust information for horizontal (binaural cues), vertical localization (elevation), and front-back discrimination, prior works indicate that inter-individual variability can introduce systematic errors such as front-back confusions or a broad cone of confusion~\cite{blauert1997spatial, huang1998spatial}. Thus, precise interpretation of spatial audio depends not only on the acoustic signal itself, but also on how a listener learns to associate auditory cues with spatial patterns. Studies show that auditory spatial perception is not fixed, but can be calibrated through experience, allowing the listener to adapt the interpretation of acoustic cues and improve spatial consistency over time~\cite{king2009visual, hrtfisgoodenough}. Developmental and perceptual research further indicates that humans gradually acquire sensitivity to spatial cues through interaction with the environment, forming expectations about how sounds correspond to locations and events~\cite{blauert1997spatial, king2009visual}. In interactive XR systems, this process can be utilized through short-term exposure and feedback, enabling users to adapt to non-individualized HRTF cue mappings, without eliminating underlying perceptual ambiguities. Visual context plays a particular role in shaping auditory spatial learning. A study demonstrated that visual signals can recalibrate perceived auditory space, biasing or correcting auditory localization through repeated co-occurrence~\cite{king2009visual, kyto2015ventriloquist, bruns2019ventriloquist}. Such cross-modal calibration does not entirely resolve the perceptual uncertainty, but it can improve consistency and confidence in directional judgments, especially for elevation and front-back distinctions which rely on the weaker directional cue (spectral).

We design our study with short-term learning as a complementary factor that shapes how spatial audio is interpreted under time-constrained conditions. By evaluating the localization performance before and after a calibration phase, we report how users adapt their interpretation of spatial audio cues, and share our findings for immediate attention guidance in XR.


\begin{figure}[!t]
    \centering
    \includegraphics[width=\linewidth]{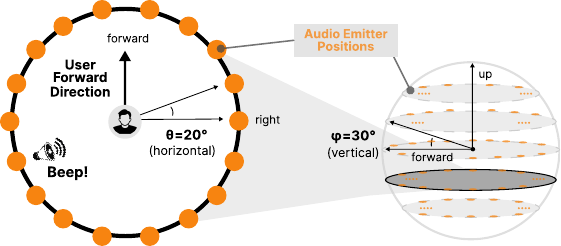}
    \caption{Layout of the audio emitters. 90 virtual sound sources are positioned on the surface of a sphere centered at the participant (r=5~meters) with a horizontal ($\theta$;  azimuth) sampling interval of $20^\circ$ over $0^\circ$ to $360^\circ$, and vertical ($\phi$, elevation) increments of $30^\circ$ over $-60^\circ$ to $60^\circ$. The top-down view (left circle) shows the horizontal ring relative to the participant's forward direction, and the side view (right sphere) illustrates the elevation rings and the spherical coordinate definition of $\theta$ and $\phi$.}
    \label{fig:experiment_settings_illustration}
\vspace{-3mm}
\end{figure}

\section{Design of Preliminary Evaluation}
Auditory notification can be used to \textit{quickly nudge} the user about imminent information, \textit{providing awareness} of any situation that requires attention. In particular, it can be valuable in industrial XR or in daily wayfinding scenarios where a user may encounter a hazard or a threat to their safety (\cref{fig:teaser}).

Auditory cues have traditionally been used to signal the occurrence of an event (``when''), while visual cues convey precise spatial information (``where''). Recent works, however, suggest that spatialized audio can partially bridge this divide by conveying coarse spatial information, enabling users to approximate the direction of a target even in the absence of visual cues. Building on these insights, we investigate whether spatial audio can serve as an effective mechanism for rapid attention capture, protecting the safety of the user in practical XR applications.

\subsection{Design Considerations}
For time-critical XR scenarios (e.g., industry hazards, an approaching motorcycle towards the user), rapidly capturing the user's attention is key. Even with a short exposure to an attention-grabbing mechanism, a user must be able to localize the threat. However, the long-standing approach to sound-source localization has been to gradually refine the perceived aural cues with head-rotation over an extended period of time~\cite{wallach1940role, houtgast1994stimulus}. We investigate the capabilities of auditory cues to explore its use-case for rapid attention capture, without these requirements.

\para{Fixed Head Orientation without Visual Cues}
To isolate human auditory spatial interpretation capabilities, we constrain head orientation and eliminate visual input during sound presentation. Constraining head orientation prevents head-driven refinement of spatial cues. Removing visual cues eliminates cross-modal influences on spatial judgment. We adopt this constraint because our target use case is the first moment of an alert, before users can reliably rotate to ``scan'' the sound. Also, fixing head orientation during audio playback prevents active-sensing strategies (e.g., micro-rotations that resolve cone-of-confusion), allowing us to quantify a lower bound for one-shot audio-based attention redirection.

\para{Audio Rendering Model}
We employ HRTF-based spatial audio rendering to approximate sound propagation from a source to a listener's ears. Although individual anatomical differences affect elevation and front-back perception, prior work shows that generic HRTFs are sufficient for conveying coarse information~\cite{hrtfisgoodenough}.

\para{Stimulus Design}
To maximize the availability of spatial cues, we employ broadband auditory stimuli with frequency ranges that maximize ITD, ILD, and spectral cues. To prevent learning biases arising from any associations between sound signals and ground-truth locations, answer feedback is withheld during localization trials.

\para{Learning and Calibration}
Auditory spatial perception can be calibrated through learning and feedback. Prior research shows that short-term exposure to aligned audio-visual cues can improve the consistency of spatial judgments, even when inherent perceptual ambiguities remain~\cite{king2009visual, kyto2015ventriloquist, bruns2019ventriloquist}. Therefore, we integrate a calibration phase within our study, and examine how the adaptation influences the localization performance.

\subsection{Experimental Design and Variables}
We employed a within-subject design comparing localization performance before and after a visuo-auditory calibration phase. The study consists of three sessions following an initial tutorial: \textbf{(Session 1)} Pre-calibration session; \textbf{(Session 2)} Calibration/Learning; and \textbf{(Session 3)} Post-calibration session. The primary independent variables were: (1) emitter direction, and (2) calibration condition (Pre vs. Post, separated by the Calibration session). Dependent variables were angular localization errors (\hori, \verti, \dist), directional confidence, and post-study confusion reports (\cref{subsec:measurements}). To quantify a lower bound for immediate, one-shot attention guidance, we removed visual landmarks during stimulus playback and constrained head orientation during playback to prevent head-driven cue refinement (\cref{subsec:task}). Each emitter directions was presented once per session, and trial order was randomized (\cref{subsec:procedure}). The deviation analyses were performed using within-subject tests (Friedman and Wilcoxon signed-rank tests with Bonferroni correction).

\subsection{Hypotheses}
In time-critical XR scenarios, users must interpret auditory cues from brief exposure, without relying on head movement or sustained listening, which require longer audio playback to judge direction. Based on our design considerations and insights derived from prior works, we formulate the following hypotheses:

\begin{enumerate}[label=\textbf{H\arabic*.}, leftmargin=*, align=left]
\item \parac{Immediate Localization Feasibility}
Spatialized audio supports approximate directional localization under brief, time-constrained playback, with performance above a permutation-derived baseline (`\textit{chance}').
\item \parac{Direction-dependent Ambiguity}
Immediate localization performance will differ by region. The confusion derived from the Left-right pair will be less than that from Front-back or Up-down.
\item \parac{Effect of Short-Term Calibration} 
Short exposure to visuo-auditory feedback will re-wire (calibrate) the human aural perception, and improve localization performance and enhance user confidence in coarse directional judgments.
\vspace{-2.5mm}
\end{enumerate}

\begin{figure}[!t]
    \centering
    \includegraphics[width=\linewidth]{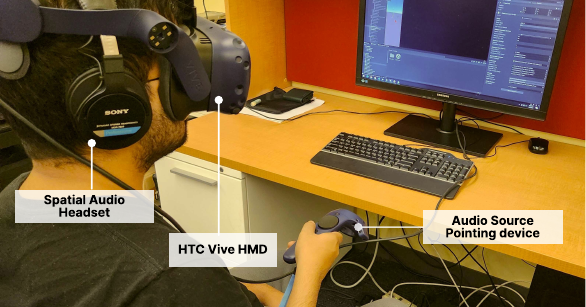}
    \caption{User study setup. Participants are instructed to point towards the perceived direction of an emitter after each stimulus.}
    \label{fig:user_study_looks_capture}
\vspace{-3mm}
\end{figure}

\subsection{Participants}
We recruited 17 participants (P1-17; 11 male, 6 female), aged 21-36 ($\mu=25.4$, $\sigma=3.9$) with prior experience with XR HMD ($\mu=2.6$, $\sigma=1.5$; 7-point Likert scale), and self-reported sensibility to sound localization ($\mu=6.3$, $\sigma=0.7$; 7-point Likert scale). No participants reported any hearing or vision deficiency. They were compensated \$10 for an estimated participation of 40 minutes. The identity of the study participants was anonymized (alias were assigned), and each participant provided informed consent prior to the participation. This study was conducted under the IRB approval of Stony Brook University (\textit{1173920}).

\subsection{Apparatus and Setup}
We conduct our experiment using an HTC Vive Pro HMD in VR settings, to avoid the introduction of uncontrolled factors such as coordinate drift or spatial misalignment that can arise in AR or MR. A wired Sony MDR-7506 headset was used for spatial audio. As illustrated in \cref{fig:experiment_settings_illustration}, we place 90 spatial audio emitters at a fixed radial distance of 5 meters ($r=5$) around the participant's center position, incrementing evenly with a horizontal interval of $20^\circ$, and a vertical interval of $30^\circ$. We cover the emitters in all horizontal angles from $0^\circ$ to $360^\circ$, and $-60^\circ$ to $60^\circ$ vertically, as illustrated in \cref{fig:experiment_settings_illustration}. We use the Steam Audio plugin in Unity to render the HRTF spatial audio. A single broadband stimuli with three second bursts of white noise, amplitude modulated at 4 Hz for improved ITD sensitivity, and localization accuracy~\cite{valzolgher2020impact, middlebrooks2015sound, houtgast1994stimulus, cho2024auptimize} was generated prior to the study, and used. The stimulus was limited to the range of 500 to 9,000 Hz, to include the strongly weighted ITD, ILD, and spectral notch regions while excluding extreme low and high frequencies that contribute relatively little spatial cue value and are perceived less reliably (non-monotonically and erratically)~\cite{zonooz, zonooz2018learning}. 

\subsection{Measurements}
\label{subsec:measurements}
\para{Angular Deviation}
Angular deviation is defined as the absolute angular difference between the ground-truth emitter direction and the participant's response direction. Localization error is decomposed into horizontal and vertical components to reflect the distinct perceptual cues involved in spatial hearing. (1) Horizontal deviation is computed as the absolute difference in azimuth angle (\hori), representing left-right localization accuracy primarily supported by binaural cues. (2) Vertical deviation is computed as the absolute difference in elevation angle (\verti), representing up-down localization accuracy primarily supported by monaural spectral cues. Then, we use (3) 3D angular distance (\dist) which encodes both horizontal and vertical at once, as the primary localization error metric. All measurements are defined in a spherical coordinate system with a fixed radial distance ($r=5$) and are reported in degrees. 

\para{Directional Confidence}
Directional confidence is collected after each, Pre-calibration and Post-calibration session using a 7-point Likert scale (1: not confident at all, 7: very confident). This self-reported metric captures the participants' perceived reliability of immediate spatial judgments, and is compared with the measured (objective) effects to identify any alignment between the results.

\para{Perceptual Confusion}
Perceived sources of confusion during localization are collected through a post-study questionnaire, including front-back ambiguity, up-down ambiguity, and left-right discrimination difficulty. Participants were given three pair options as multiple choice questions (allowing multiple answers). These reports help coarsely rank the perceptual difficulty of audio source localization, and to compare any alignment between self-reports and the measured effects.

\subsection{Task}
\label{subsec:task}
In each trial, a single broadband spatial audio stimulus is presented from a pre-defined direction around the participant (\cref{fig:experiment_settings_illustration}). The participants are asked to infer the perceived direction of the sound source immediately after the stimulus playback. Responses were provided via a pointing-based interaction (\cref{fig:user_study_looks_capture}). The task was repeated both before and after a calibration/training phase, allowing localization performance to be compared under identical task conditions.

\subsection{Procedure}
\label{subsec:procedure}
The participants first received an overview of the study tasks and instructions, followed by a 10-minute tutorial to familiarize themselves with the study settings. During Sessions 2 and 3, participants were instructed to maintain a forward-facing head orientation until the completion of an auditory stimulus in each trial, preventing head-driven refinement of spatial cues, during listening. No additional audio was played after this stage. The virtual environment during stimulus playback consists of a black scene with no visual landmarks, eliminating any visual cues that could bias the auditory interpretation. 

After the audio stimulus was presented from a pre-defined direction, participants were allowed to rotate or move, and were asked to infer the perceived direction of the sound source, and indicate their response by pointing towards the inferred location using a handheld controller, and confirm their selection with a trigger button. The ray-based pointing technique--casting a ray towards an invisible spherical bounds where emitters are situated--was used to select the inferred source location. No visual representations of the sound emitters were provided, to exclude any visual-bias in the inference. \cref{fig:user_study_looks_capture} illustrates our experimental setup and an example participant interaction. The Pre-calibration and Post-calibration sessions used identical task settings and configurations. Between these two sessions, participants completed a Calibration session in which spatialized auditory stimuli were paired with visual indicators of the ground-truth source location. Its configuration is identical to other sessions, but with added visual feedback. This feedback design was intended to allow the participants to calibrate their aural perception, finding the correspondence between aural patterns and spatial locations. To fairly compare performance across emitters and reduce learning effects (latter trials being more accurate), participants were not provided with feedback on their inference except during Calibration. The three sessions use identical configurations and each session presents the full set of emitter directions once, with a balanced random order.

\section{Results} 
A total of 4,590 localization trials were collected (17 participants x 3 sessions x 90 trials), each session contributing 1,530 trials. Tutorial trials were logged as well, but were excluded from our analyses. 

\begin{figure*}[!t]
    \centering
    \includegraphics[width=\linewidth]{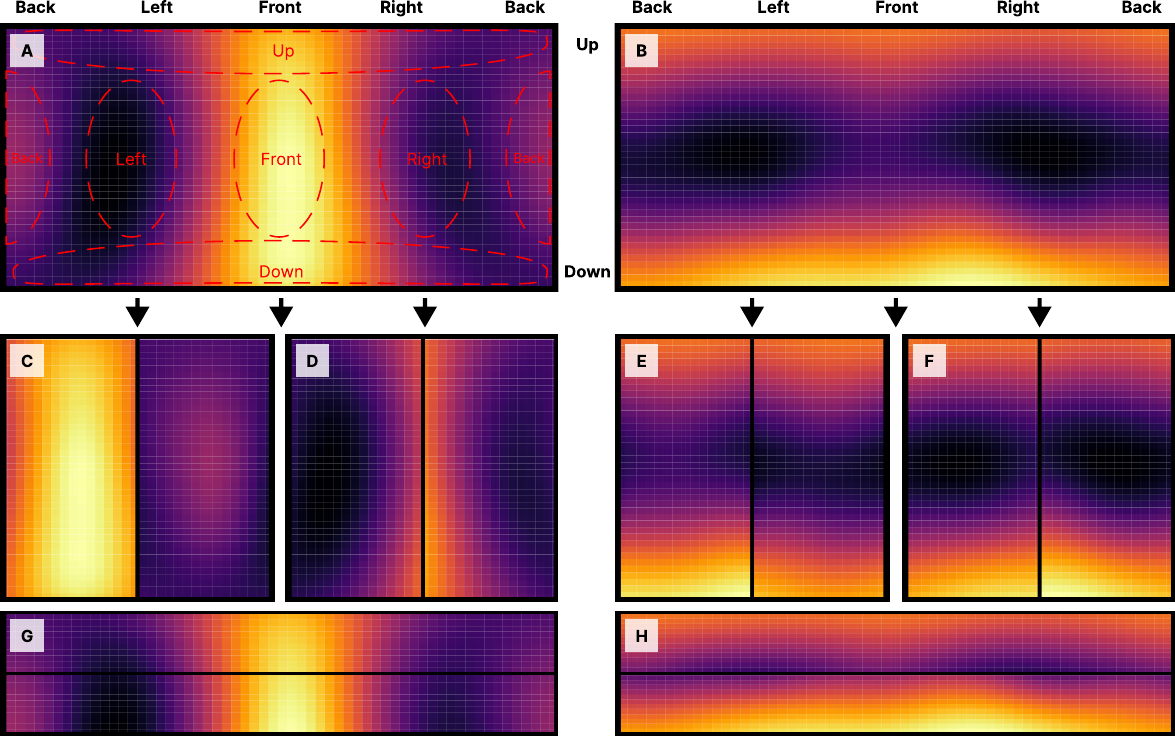}
    \caption{2D projection-mapped visualization of localization error of each axis (horizontal/vertical) across directions. Emitter directions on the sphere are mapped to an equirectangular 2D grid (horizontal axis: labeled Back-Left-Front-Right-Back; vertical axis: labeled Up-Down). \textcolor{heatmap_yellow}{\textbf{Brighter colors}} indicate \uline{larger errors} and \textcolor{heatmap_purple}{\textbf{Darker colors}} for \uline{low error regions}. \textbf{(A, B)} Horizontal ($\theta$) and Vertical ($\phi$) deviation visualization (ground-truth to user-inference error). Note the markings (in \textcolor{red}{red dashed line}) of each region in \textbf{(A)}; \textbf{(C,D,E,F,H)} pairs its opposite direction-pair to visualize the side-by-side comparison of the error rate visualization. \textbf{(C,E)} indicate Front-Back, \textbf{(D,F)} depict the Left-Right, and \textbf{(G,H)} show Up-Bottom pairs, each representing the segmented sub-components of \textbf{(A)} and \textbf{(B)} respectively. The visualization in \textbf{(A)} depicts the error-prone emitter regions for horizontal dimension-azimuth ($\theta$), and \textbf{(B)} for vertical-elevation ($\phi$)}
    \label{fig:heatmap}
\vspace{-3mm}
\end{figure*}

\para{Localization Accuracy}
\label{para:localization_accuracy}
The localization performance under a short broadband audio exposure, was noisy across the sessions. The mean absolute errors of \pre were: \hori$=61.14^\circ$, \verti$=38.97.14^\circ$, \dist$=69.19^\circ$, and for \post: \hori$=57.14^\circ$, \verti$=38.00^\circ$, \dist$=65.38^\circ$. To evaluate whether the performance could occur by chance, we conducted a permutation test (randomly pairing responses from 1,000 samples) to build a baseline. It confirmed that our result did not happen by chance. The mean permutation baseline was less accurate than our reports. The \pre result of baseline indicates \hori$=90.05^\circ$, \verti$=41.85^\circ$, \dist$=89.97^\circ$ (${p<0.001}$ across all metrics; \post was significance as well). 

We assess whether spatial audio can serve as a coarse attention-capturing cue by measuring the proportion of trials whose response direction falls within an angular tolerance around the ground-truth direction. Using the 3D angular error, \dist (the separation angle between the response and target vectors), we count ``successful'' trials within cone half-angles of $45^\circ$, $60^\circ$, and $90^\circ$, which translates to the full cone/FoV angles of $90^\circ$, $120^\circ$, and $180^\circ$ (hemisphere of a sphere). This use of a cone is a mock frustum FoV for XR. In \pre, $27.65\%$ of the trials fell within $45^\circ$ from the ground-truth emitter position, $44.18\%$ were within $60^\circ$, and $74.51\%$ within $90^\circ$. In \post, these increase to $33.01\%$ ($45^\circ$),  $49.08\%$ ($60^\circ$), and $74.64\%$ ($90^\circ$).

\para{Direction-dependent Ambiguity}
We group the sound emitters into 6 high-level directional regions (Front, Back, Left, Right, Up, and Down) to observe whether the well-established confusion patterns (Front-back confusion and Up-down confusion) recur in our scenario where there is no head rotation-based aural cue refinements. We define the 24 closest sampled emitters to the Cartesian axis direction ($\pm$X, $\pm$Y, $\pm$Z axes) from the user's front-facing direction, as the six directional region.

In \pre, the mean 3D angular distance by region was: Front=$93.29^\circ$ ($\pm36.2^\circ$), Back=$54.50^\circ$ ($\pm32.9^\circ$), Left=$58.62^\circ$ ($\pm30.0^\circ$), Right=$62.34^\circ$ ($\pm32.8^\circ$), Up=$66.69^\circ$ ($\pm31.62^\circ$), and $75.65^\circ$ ($\pm35.1^\circ$) for Down. The mean 3D angular distance for \post was: Front=$82.09^\circ$ ($\pm37.1^\circ$), Back=$60.50^\circ$ ($\pm36.9^\circ$), Left=$53.19^\circ$ ($\pm30.0^\circ$), Right=$61.51^\circ$ ($\pm34.3^\circ$), Up=$62.33^\circ$ ($\pm33.1^\circ$), and $69.95^\circ$ ($\pm35.7^\circ$) for Down.

The Friedman test for \pre, confirmed this regional ambiguity ($\chi^2=36.66, p<0.001)$. Out of the 15 pairs, 8 regions were found significant. Larger values indicate larger deviation (more error-prone). 
$Front{>}Left$ $_{(\Delta=34.67^\circ; p{<}0.001)}$,
$Front{>}Right$ $_{(\Delta=30.95^\circ; p{<}0.01)}$,
$Front{>}Back$ $_{(\Delta=38.78^\circ; p{<}0.01)}$,
$Front{>}Up$ $_{(\Delta=26.60^\circ; p{<}0.01)}$,
$Left{<}Down$ $_{(\Delta=17.04^\circ; p{<}0.01)}$,
$Right{<}Down$ $_{(\Delta=13.32^\circ; p\approx0.03)}$,
$Front{>}Down$ $_{(\Delta=17.63^\circ; p\approx0.04)}$, and 
$Back{<}Down$ $_{(\Delta=21.15^\circ; p{<}0.001)}$. The mean 3D error of opposite direction-pair comparisons (e.g., Front-back, Up-down) showed significance for $Front_{vs}Back$ (Wilcoxon, $p{<}0.001$). The other pairs, $Left_{vs}Right$ and $Up_{vs}Down$ showed no significance. We also quantify the opposite-pairs using the percentage of trials to outline the confusion rates. The confusion rate of $Front{-}Back$ was $49.14\%$, $Left{-}Right$=$7.23\%$, and $Up{-}Down$=$43.30\%$. Significance differences were found for $Pair_{(Left{-}Right)}{<}Pair_{(Front{-}Back)}$ $_{(\Delta=41.91\%; p{<}0.001)}$, $Pair_{(Left{-}Right)}{<}Pair_{(Up{-}Down)}$ $_{(\Delta=36.07\%; p{<}0.001)}$, and $Pair_{(Front{-}Back)}{>}Pair_{(Up{-}Down)}$ $_{(\Delta=5.84\%; p\approx0.04)}$.

All in all, our result indicates that the Front region has the highest audio localization error, and participants were not able to distinguish between Front and Back. This result also aligned with the post-study qualitative feedback: ``\textit{Front and Back I can't tell the difference much.}'' (P17) and ``\textit{Are you sure the sound is really coming from the Front not Back?}'' (P2). For the amount of confusion from the participants, $Pair_{(Front{-}Back)}$ was highest, followed by $Pair_{(Up{-}Down)}$, and $Pair_{(Left{-}Right)}$ being the lowest. This also closely aligned with the frequency of participants' self-reported ambiguity pair (higher means more difficult to distinguish; $Pair_{(Front{-}Back)}$=11; $Pair_{(Up{-}Down)}$=11; $Pair_{(Left{-}Right)}$=0). The visualization of each pair is illustrated in \cref{fig:heatmap}.

\para{Effects of Short Term Calibration}
As reported in \hyperref[para:localization_accuracy]{Localization Accuracy}, the overall mean angular distance between \pre and \post ($\Delta_{pre,post}$) reduced by $3.81\pm5.81^\circ$ (from $69.19^\circ\pm35.6^\circ$ to $65.38^\circ\pm36.1^\circ$; Significant, $p=0.015$). For each of the six regions, Front showed significant deviation reduction ($\Delta=11.9^\circ$; $p=0.04$) after the Calibration session. The Left also showed significance in error reduction ($\Delta=5.43^\circ$; $p=0.03$). However, although the other regions (Up, Down, Right) showed reduced \dist, they did not exhibit statistical significance. Back, however, had an increase in error rate ($\Delta=11.19^\circ$, no significance) after calibration. 

The short-term exposure to aural perception calibration reduced the opposite-pair confusion, but was not effective enough (no pair reached significance). The confusion between left and right $Pair_{(Left{-}Right)}$ reduced from $7.23\%$ to $6.37\%$ ($\Delta=0.86\%$), the $Pair_{(Front{-}Back)}$ reduced from $49.14\%$ to $46.49\%$ ($\Delta=2.45\%$), and the confusion between up and down $Pair_{(Up{-}Down)}$ reduced from $43.30\%$ to $39.95\%$ ($\Delta=3.35\%$).

\begin{table}[!t]
\centering
\caption{Summary statistics of mean 3D angular localization error (\dist, in degrees), by overall and regions before (Pre) and after (Post) calibration; $\Delta$ denotes the shift in mean error (Post-to-Pre calibration).}
\label{tab:region_stats}
\small
\begin{tabular}{|p{0.18\columnwidth}|c|c|c|c|c|}
\hline
\textbf{Region} & \multicolumn{2}{c|}{\textbf{Pre}} & \multicolumn{2}{c|}{\textbf{Post}} & \textbf{$|\Delta|$ Mean} \\
\cline{2-5}
 & \textbf{Mean} & \textbf{Std} & \textbf{Mean} & \textbf{Std} & \textbf{(Post--Pre)} \\
\hline
Overall & $69.19^\circ$ & $35.6^\circ$  & $65.38^\circ$ & $36.1^\circ$ & $3.81^\circ$ \\
Front   & $93.29^\circ$ & $36.2^\circ$  & $82.09^\circ$ & $37.1^\circ$ & $11.20^\circ$ \\
Back    & $54.50^\circ$ & $32.9^\circ$  & $60.50^\circ$ & $36.9^\circ$ & $6.00^\circ$ \\
Left    & $58.62^\circ$ & $30.0^\circ$  & $53.19^\circ$ & $30.0^\circ$ & $5.43^\circ$ \\
Right   & $62.34^\circ$ & $32.8^\circ$  & $61.51^\circ$ & $34.3^\circ$ & $0.83^\circ$ \\
Up      & $66.69^\circ$ & $31.62^\circ$ & $62.33^\circ$ & $33.1^\circ$ & $4.36^\circ$ \\
Down    & $75.65^\circ$ & $35.1^\circ$  & $69.95^\circ$ & $35.7^\circ$ & $5.70^\circ$ \\
\hline
\end{tabular}
\vspace{-3mm}
\end{table}

\section{Summary and Insights} 
We evaluate whether a brief spatial audio cue can function as an immediate, coarse attention-guidance mechanism for XR in time-critical scenarios (e.g., hazard or collision avoidance alerts), where users cannot rely on extended listening or head-driven refinement. To quantify the intrinsic capability of spatial audio as a rapid ``attention nudge,'' we isolate the auditory cue and remove any visual landmarks, measuring a lower bound on aural performance.

Overall, our results support that spatial audio is a viable solution as a coarse orienting cue, but not as a precision pointer under immediate use. The participants' inferred direction yielded meaningful results (better than `chance'), indicating that short spatialized audio can bias a user's attention towards the target region, and guide the user for initial reorientation (summary of localization error: \cref{tab:region_stats}) -- \textbf{(RQ1, H1)}. However, the remaining localization uncertainty is large (Post-calibration mean: \dist=$65.38^\circ$), suggesting that audio alone is insufficient when the application demands high directional precision.

We further validate that immediate localization is strongly direction-dependent \textbf{(RQ2, H2)}, and highlight the importance of gradual cue refinement (head reorientation). The lateral judgments (left-right) were comparatively stable, while front-back and vertical distinctions were substantially ambiguous. This pattern is consistent with the perceptual limits implied by the ``cone-of-confusion'' and the weaker reliability of spectral cues under a generic HRTF rendering model (given anatomical differences across listeners)~\cite{waspaa}. Thus, a single-shot spatial audio cue is the most robust for Left-Right orienting, but must be carefully controlled for front-back or up-down disambiguation, unless additional cues are provided.

We validate that short-term visuo-auditory remapping can yield measurable, but bounded, benefits even under non-head-refined use of audio \textbf{(RQ3, H3)}. While overall directional inference improved after calibration/feedback, the dominant ambiguity patterns were not eliminated, implying that brief training can tune users' internal mapping to the rendered spatial cues, but may not fully overcome inherent confusions in time-constrained, one-shot settings.

Our findings imply several preliminary design guidelines for time-critical XR notification designs. (1) First, spatial audio should be positioned as a first-stage attention signal: it can rapidly attract attention and prompt coarse reorientation, but it is not recommended as the sole mechanism when precise ``where'' information is required, especially with the use of a generic HRTF model. (2) Second, an interface may need to explicitly handle front-back and up-down ambiguity (e.g., pairing spatial audio with other multimodal signals), or extend adaptive cue transformation solutions to reduce confusable configurations~\cite{cho2024auptimize}. (3) Finally, the results suggest diverse future research directions: a richer visual context, personalized HRTFs, or perceptually-informed audio transformations (particularly for front-back confusions).


\section{Future Work and Discussion}
Building on the insights learned from this study--the potential and limitation of spatial audio as a coarse attention redirection mechanism, we plan to incorporate motion and visual context in future work to better comprehend the reliability of the use of audio-based cues for safety-critical notifications.

\para{Head-motion as continuous refinement}
A key open question is whether localization improves linearly when head-motion is allowed, and whether the head rotation can be modeled as an aggregated sequence of short, single-frame inferences rather than a one-time estimate. In our follow-up work, we will quantify how accuracy changes as a function of rotation amount, speed, and temporal integration (e.g., whether a brief cue repeated across micro-rotations yields any predictable convergence). This reframes localization as an active sensing process and may inform how XR systems should adapt alerts for rapid attention guidance.

\para{Visual context and front-back disambiguation}
Our results show that front-back ambiguity remains a dominant failure mode under an immediate, audio-only condition. In realistic XR scenarios, the forward visual field is often the primary source of spatial grounding, and audio may function as an auxiliary trigger to acquire the relevant visual evidence. We plan to expand the study by introducing controlled visual context (e.g., sparse landmarks, minimal forward FoV previews) to quantify how much vision resolves front-back errors, and whether the improvement is asymmetric for frontal objects. This can also reveal whether the poor performance for frontal sources reflects a perceptual limitation of the rendered audio, or a reliance on cross-modal consistency which was unavailable under our one-shot settings.

\para{Competing task and performance}
We isolate auditory localization to quantify a baseline for one-shot spatial cues. However, in real XR scenarios, alerts may arrive while users are visually engaged in a primary task (e.g., inspection, navigation, or manipulation), and divided attention may increase localization error and response time. As our future work, we plan to conduct a dual-task study~\cite{kaur2025senses} that compares the performance differences between our audio-only baseline and task-occupied, and measures the localization error, reaction time, perceived cognitive load, and interference.

\para{Beyond a single-time calibration}
While the short-term visuo-auditory feedback improved overall directional inference accuracy, it did not eliminate the dominant confusion patterns. In our future work, we will treat calibration as a dynamic, interactive process rather than a discrete phase by varying feedback frequency and timing. This may reveal whether an XR system should include a continuous calibration phase based on the disparity between predicted and user input.

\para{Varying performance per audio engine}
We utilize Steam Audio to generate head-related spatial audio (generic HRTF), but as the localization performance may vary across different audio pipelines and HRTF engines, we plan to extend our work to alternative spatializers (e.g., Dolby Atmos via Wwise or FMOD), and observe differences in perceptual ambiguity. 

\para{Perceptually informed stimulus shaping}
We use a single broadband cue to maximize access to ITD, ILD, and spectral information, but the contribution of different frequency regions is not uniform. Thus, we plan to research perceptually informed stimulus design, including spectral-weighted approaches~\cite{zonooz, zonooz2018learning} that weigh informative bands while down-weighting less reliable extremes, as well as temporal shaping (burst length, onset/offset, modulation), to explore any improvement in aural perception performance.

\section{Conclusion} %
We investigated whether spatial audio can serve as an immediate attention-capturing cue in time-critical XR when users cannot rely on extended listening or head-driven refinement. We quantified a lower bound of spatial audio localization performance, by removing visual stimuli and measuring localization errors. We showed that brief HRTF-rendered cues can convey coarse directional information, but ambiguities remain, with strongly direction-dependent errors. Short-term exposure to visuo-auditory feedback yields measurable performance improvements, indicating that quick calibration can tune users' mapping to an audio renderer, but does not eliminate the inherent limitations. We empirically assess spatial audio as a rapid first-stage attention cue in XR, clarify the perceptual boundaries of immediate interpretation, and provide preliminary design guidelines along with future work directions for time/safety-critical user attention capturing.

\acknowledgments
{
\cref{fig:teaser} was edited using ChatGPT. This work was supported in part by NSF awards IIS2529207 and ONR award N000142312124.
}

\bibliographystyle{abbrv-doi-hyperref-narrow}

\bibliography{references}
\end{document}